\documentclass[1,12pt]{elsarticle}

\usepackage{graphicx}
\usepackage{subfigure}

\usepackage{amssymb,amsmath}
\usepackage{exscale}
\usepackage{makeidx,shortvrb,latexsym}
\usepackage{epstopdf}
\usepackage{tabularx, booktabs, multirow}
\usepackage[]{hyperref}
\newcommand{\rd}{\mathrm{d}}

\usepackage{color}

\begin{document}

\begin{frontmatter}

\title{An analysis of the influence of grain size on the strength of FCC
polycrystals by means of computational homogenization}

\author{Sarra Haouala$^1$}
\author{Javier Segurado$^{1, 2 }$}
\author{Javier LLorca$^{1, 2, }$\corref{cor1}}
\address{$^1$IMDEA Materials Institute \\ C/ Eric Kandel 2, 28906 - Getafe, Madrid \\\ \\
$^2$Department of Materials Science, Polytechnic University of
  Madrid \\ E. T. S. de Ingenieros de Caminos. 28040 - Madrid, Spain.}
\cortext[cor1]{Corresponding author}

\begin{abstract}

The effect of grain size on the flow stress of FCC polycrystals is analyzed by means of a multiscale strategy based on computational homogenization of the polycrystal aggregate. The mechanical behavior of each crystal is given by a  dislocation-based crystal plasticity model in which the critical resolved shear stress follows the Taylor model. The generation and annihilation of dislocations in each slip system during deformation is given by the Kocks-Mecking model, which was modified to account for the dislocation storage at the grain boundaries. Polycrystalline Cu is selected to validate the simulation strategy and all the model parameters are obtained from dislocation dynamics simulations or experiments at lower length scales and the simulation results were in good agreement with experimental data in the literature. The model is applied to explore the influence of different microstructural factors (initial dislocation density, width of the grain size distribution, texture) on the grain size effect. It is found that the initial dislocation density, $\rho_i$, plays a dominant role in the magnitude of the grain size effect and that dependence of flow stress with an inverse power of grain size ($\sigma_ y -\sigma_\infty \propto d_g^{-x}$) breaks down for large initial dislocation densities ($> 10^{14}$ m$^{-2}$)  and grain sizes  $d_g > $ 40 $\mu$m in FCC metals. 
However, it was found that the grain size contribution to the strength followed a power-law  function of the dimensionless parameter $d_g\sqrt{\rho_i}$ for small values of the applied strain ($<$ 2 \%), in agreement with previous theoretical considerations for size effects in plasticity. 

\end{abstract}


\end{frontmatter}

\section{Introduction}

The bonds between metallic atoms lead to crystalline materials with high stiffness that can withstand  plastic deformations and dissipate large amounts of energy before failure. These properties are ideal for structural applications but the stress necessary to promote plastic deformation is very low in most metals. Different strategies have been developed to overcome this limitation and solid-solution, precipitation and strain hardening are often combined to increase the density and strength of obstacles to the dislocation motion and to enhance the flow stress of metals and metallic alloys. Moreover, metallic alloys are often used as polycrystals and it is well established that the strength of polycrystalline metals can also be increased by reducing the grain size. The pioneer work of Hall \cite{H51} and Petch \cite{P53} established a phenomenological dependence of the yield strength, $\sigma_y$, with the grain size, $d_g$, of the form,

\begin{equation}
\sigma_y=\sigma_\infty+C_{HP}d_g^{-0.5}
\label{eq:HP}
\end{equation}

\noindent where $\sigma_\infty$ is the yield strength of a polycrystal with very large grain size and $C_{HP}$ is a material constant. Eq. (\ref{eq:HP}) was supported by the work of Eshelby {\it et al.} \cite{EFN51} for the stress necessary to move a dislocation in front of a dislocation pile-up formed at the grain boundary and also from work hardening models that assume that flow stress increases with the square root of the dislocation density \cite{H72}. Further support for eq. (\ref{eq:HP}) was provided by Ashby \cite{A70}, who analyzed the plastic incompatibility between grains with different orientation within the polycrystal. The increase in dislocation density that leads to hardening could be separated into two different contributions. Statistically stored dislocations (SSDs) account for a uniform deformation, while  geometrically necessary dislocations (GNDs) are required to preserve the lattice continuity between grains with different orientation. The density of the former is grain-size independent, while that of the latter is concentrated around the grain boundaries and depends on the grain size.

However, the generality of eq. (\ref{eq:HP}) was challenged, as many authors reported that most of the experimental data could also be fitted with $d_g^{-x}$ with $ 0 < x \le 1$ \cite{RP86}. Other authors \cite{DB13, DB14} found that the experimental data better supported $x$ = -1 or a dependence on grain size of the type $(\ln d)/d$ \cite{H72}. The former exponent was in agreement with a grain representation formed by a soft core surrounded by hard shell around the grain boundary \cite{K70, H72} while the latter was consistent with a mechanism in which the grain size constrains the size of the dislocation sources \cite{LBD16}.

Today it is acknowledged that the increase of the strength of polycrystals with grain size is a manifestation of the general size effect found in plasticity \cite{LBD16, LDH05} and the dominant mechanism(s) (and, thus the value of the exponent $x$) depends on many factors, such as the elastic anisotropy of the crystal, the range of grain sizes examined, the texture, the number of slip systems, the initial dislocation density, the presence of other obstacles to dislocation motion, etc. \cite{BDN08, FBM01}. As the specific influence of each of these factors is very difficult to be accounted for separately in experiments, numerical simulations become very useful to understand the role played by each one. In the particular case of polycrystalline aggregates, computational homogenization in combination with crystal plasticity has demonstrated its potential to simulate the effective properties of polycrystals while the details of the deformation within the grains are taken into by the crystal plasticity constitutive equation \cite{MSS93, LLP04, LRS11, SLL14}.

Several attempts can be found in the literature to simulate the effect of grain size on the mechanical behavior of polycrystals. The first attempt was due to Weng \cite{W83}, who introduced a grain-size dependent constitutive equation for the slip deformation of slip systems. However, the influence of grain size is a macroscopic result and should be an outcome and not an input of the model. Another  attempt to capture grain size effects was based on the self-consistent homogenization scheme in which each grain of the polycrystal was represented as a two-phase composite: a core region in which the strain hardening results from the evolution of SSDs and an interphase layer which corresponds to the grain boundary region, where plastic strain gradients and associated GNDs are present \cite{PNB09}. This model was successfully applied to predict the effect of grain size on the flow stress of ferritic steels with different grains sizes (in the range 5.5 $\mu$m to 120 $\mu$m) but it should be noted that the thickness of the grain boundary region was an adjustable parameter to fit the experimental data. 

Homogenization models of polycrystals based on classical plasticity cannot capture the grain size effect because the constitutive equation does not involve an intrinsic materials length scale. This limitation can be overcome by introducing a length associated  with strain gradients in continuum crystal plasticity models \cite{AB00, EPB02, CBA05, BBG07, BER10}. Hardening around the grain boundaries comes about as result of the strain gradients (and the associated density of GNDs) which arise to maintain the lattice compatibility between grains with different orientation. However,  a direct comparison of these models  with the actual hardening found in polycrystals has not been carried out and the physical origin of the length scale included in the formulation is not clear in the case of phenomenological models  although this parameter controls the magnitude of the size effect \cite{BER10, BSP13}. More recently, Wagoner and co-workers \cite{LLK11, LSF14} presented another approach that did not invoke any arbitrary length scale. Polycrystal simulations were carried out using a dislocation-based crystal plasticity model. This information was used at another scale to enforce local slip transmission criteria at the grain boundaries depending on the orientation and on the grain boundary strength. 

In this investigation, a multiscale approach is used to analyze within the framework of the computational homogenization of polycrystals. The mechanical response of each crystal follows a rate dependent physically-based crystal plasticity model in the context of finite strain plasticity. The critical resolved shear stress on each slip system in the model is linked with the dislocation densities by a Taylor model \cite{T34} in which the strengthening provided by the different types of interactions among dislocations are obtained from dislocation dynamics simulations. The evolution of dislocation density in each slip system was governed by a Kocks-Mecking law \cite{KM81, KM03} in which the term that controls the multiplication of dislocations, which is inversely proportional to the dislocation mean free-path, also takes into account  the dislocation storage at the grain boundary \cite{SDK10}. The model parameters in the case of Cu were obtained from simulations at lower length scales so the predictions of grain size strengthening in polycrystals are free of adjustable parameters. The multiscale approach was validated by comparison with experimental data in the literature  and the influence of different microstructural factors (grain size, grain size distribution, texture, initial dislocation densities, etc.) on the Hall-Petch behavior was ascertained.

The outline of the paper is the following. After the introduction, the crystal plasticity model is presented  in Section \ref{sec1} and the computational homogenization strategy in Section \ref{sec2}.  The simulation results and the corresponding comparison with experimental data are included in Section \ref{sec3}, while the main conclusions of the paper are summarized in the last section. In the following, vectors, second and fourth rank tensors are denoted by $\mathbf{a},\,\mathbf{A},\,\mathbb{A}$. A Cartesian coordinate system is used with respect to the orthonormal basis $\left(\mathbf{e}_{1},\mathbf{e}_{2},\mathbf{e}_{3}\right)$. The notations for tensor product, contraction and double contraction products are:
$ \mathbf{a}\otimes\mathbf{b}=a_ib_j\mathbf{e}_{i}\otimes\mathbf{e}_{j}; \mathbf{A}\cdot\mathbf{B=}{\sum}A_{ik}B_{kj}\left(\mathbf{e}_{i}\otimes\mathbf{e}_{j}\right) $ and
$\mathbf{A}:\mathbf{B=}A_{ij}B_{ij} $. Finally $\mathbf{1}$ and $\mathbb{I}$ stand for the second and fourth order identity tensors, respectively.

\section{Crystal plasticity model}\label{sec1}

The crystal plasticity model assumes a multiplicative decomposition of the deformation gradient $\mathbf{F}$ into elastic $\mathbf{\, F}^{e}$ and plastic $\mathbf{F}^{p}$ parts according to \cite{LL67},

\begin{equation}\label{eq1}
\mathbf{F}=\mathbf{F}^e\cdot\mathbf{F}^p
\end{equation}

\noindent where the configuration defined by $\mathbf{F}^p$ is called the relaxed or intermediate
configuration. The velocity gradient $\mathbf{L}$ can be expressed as

\begin{equation}
\mathbf{L}=\dot{\mathbf{F}}\cdot \mathbf{F}^{-1}=\mathbf{L}^e+\mathbf{F}^e\cdot\mathbf{L}^p\cdot\mathbf{F}^{e^{-1}}
\end{equation}

\noindent where the superposed dot denotes the total derivative with respect to time and $\mathbf{L}^e$ and $\mathbf{L}^p$ are defined as

\begin{equation}
\mathbf{L}^e= \dot{\mathbf{F}}^e\cdot\mathbf{F}^{e^{-1}}, \quad \mathbf{L}^p= \dot{\mathbf{F}}^p\cdot\mathbf{F}^{p^{-1}}.
\end{equation}

Plastic deformation in the single crystal takes place along different slip systems $\alpha$ where $n$ is the total number of slip systems. The crystallographic split on the plastic flow rate is given by

\begin{equation}
\mathbf{L}^p=\sum_{\alpha} \dot{\gamma}^{\alpha}\left(\mathbf{s}^{\alpha} \otimes \mathbf{m}^{\alpha} \right),
\end{equation}

\noindent where $\dot{\gamma}^{\alpha}$ stands  for the plastic shear strain on the slip system $\alpha$ and 
$\mathbf{s}^{\alpha}$ and $\mathbf{m}^{\alpha}$ denote, respectively, the unit vectors in the slip direction and perpendicular to the slip plane normal in the intermediate configuration.

The second Piola-Kirchhoff stress tensor $\mathbf{S}$ is expressed in terms of the elastic Green-Lagrange strain tensor $\mathbf{E}^e$, both relative to the intermediate configuration, as

\begin{equation}
\mathbf{S}=\mathbb{C}:\mathbf{E}^e \qquad  \text{and} \qquad \mathbf{E}^e=\frac{1}{2}\left(\mathbf{F}^{e^{T}}\cdot\mathbf{F}^{e}- \mathbf{1}\right),
\end{equation}

\noindent where $\mathbf{\mathbf{\mathit{\mathbb{C}}}}$ is the elastic stiffness tensor of the crystal. The resolved shear stress, $\tau^{\alpha}$, can be defined as the projection of the Piola-Kirchhoff stress on the corresponding slip system, and it is given in the intermediate configuration by

\begin{equation}
\tau^{\alpha}=\mathbf{S}:\left(\mathbf{s}^{\alpha} \otimes \mathbf{m}^{\alpha} \right).
\end{equation}

Finally, the Cauchy stress $\boldsymbol\sigma$ can be obtained as

\begin{equation}
\boldsymbol\sigma= J^{-1}\mathbf{F}^e\cdot\mathbf{S}\cdot\mathbf{F}^{e^T}, \: \text{with} \; J=det(\mathbf{F}).
\end{equation}

The relationship between the  resolved shear stress in the slip system $\alpha$, $\tau^{\alpha}$,  and the corresponding plastic strain rate, $\dot{\gamma}^{\alpha}$, is given by dislocation theory according to  (\cite{K75, KL78})
\begin{equation}
\dot{\gamma}^{\alpha}=\dot{\gamma}_0\left(\frac{|\tau^{\alpha}|}{\tau^{\alpha}_c}\right)^{m} sgn(\tau^{\alpha}),
\end{equation}

\noindent where $m$ is the strain-rate sensitivity coefficient, $\dot{\gamma}_{0}$ the reference shear strain rate and $\tau_{c}^{\alpha}$ the critical resolved shear stress on the slip system $\alpha$. 

Physically-based hardening models assume that the CRSS is proportional to the dislocation density \cite{T34}. This relationship was generalized by Franciosi {\it et al.} \cite{FBZ80} to account for the anisotropy of the interactions between different slip systems according to

\begin{equation}\label{tauc}
\tau^{\alpha}_c= \mu b\sqrt{\sum_{\beta}a^{\alpha \beta} \rho^{\beta}},
\end{equation}

\noindent where $\mu$ and $b$ denote the shear modulus and the Burgers vector, respectively, and $\rho^\beta$ stands for the dislocation density in the slip system $\beta$. The dimensionless coefficients $a^{\alpha\beta}$ of the dislocation interaction matrix represent the average strength of the interactions between dislocations in pairs of slip systems.  Recent 3D dislocation dynamics simulations \citep{E15}, carried out in cylindrical single crystals with a diameter $D$ in the range 0.25 $\mu$m $\le D \le 20 \le \mu$m, have shown that the traditional Taylor model in eq. (\ref{tauc}) should be modified by adding another term of the form $\beta \mu /(D\sqrt{\rho})$, where $\beta$ = 1.76 $\times$ 10$^{-3}$ is a constant  and $D$ the diameter of the cylinder. This new term accounts for the strength of the weakest dislocation source in the crystal and it is relevant in the case of small crystals with low dislocation densities. In the range of crystal sizes ($>$ 10 $\mu$m) and dislocation densities ($>$ 10$^{12}$ m$^{-2}$) analyzed in this investigation, the magnitude of this hardening contribution is negligible and, thus, this term was not included in eq. (\ref{tauc}).

FCC crystals have 12 $\{111\}<110>$ slip systems but only six independent coefficients are necessary to determine the 12 $\times$ 12 coefficients of the interaction matrix due to symmetry considerations \cite{DHK08}. Three of them account for different types of forest interactions between dislocations: self-interaction of dislocations in the same slip system (same slip plane and Burgers vector), coplanar dislocations (same slip plane but different Burgers vector) and collinear interaction (dislocations on different planes with the same Burgers vector). The remaining three coefficients stand for the effect of dislocation junctions in FCC crystals. They include the formation of glissile junctions between coplanar dislocations with different Burgers vector (leading to a  glissile dislocation), the Hirth lock formed by the intersection between two perfect dislocations with non-coplanar Burgers vectors that glide on intersecting planes and the Lomer-Cottrell lock that develop between Shockley partial dislocations on two intersecting \{111\} planes \cite{DHK08, BCB13}. The magnitude of interaction coefficients for different types of interactions in various lattices (FCC, HCP, BCC) can be determined by means of discrete dislocation dynamics simulations \cite{DKH06, QMD09, BTB14}. In the particular case of FCC crystals, they were obtained in \cite{DHK08, BCB13} and can be found in Table \ref{tab1}.

The overall hardening of the crystal during deformation is controlled by the evolution of the dislocation density. According to Kocks and Mecking \cite{KM81, KM03} and Teodosiu \cite{T97}, the accumulation rate of dislocations in each slip system $\alpha$,  $\dot{\rho}^{\alpha}$, can be expressed as

\begin{equation}\label{evolution-dislocation}
\dot{\rho}^{\alpha}=\frac{1}{b} \left(\frac{1}{\ell^{\alpha}}- 2y_c \rho^{\alpha} \right)|\dot{\gamma}^{\alpha} |.
\end{equation} 

\noindent The first term within the parenthesis expresses the dislocation accumulation rate and depends on the dislocation Mean Free Path (MFP), $\ell^{\alpha}$, which stands for the distance travelled for a dislocation segment before it is stopped by an obstacle. The second term within the parenthesis stands for the dislocation annihilation due dynamic recovery  and depends on the actual dislocation density $\rho^{\alpha}$ and on $y_c$, which stands for the critical annihilation distance for dislocations. 
This annihilation distance depends on the type of dislocation (either edge or screw) and on the deformation regime. Experimental observations in Cu single crystals \citep{EM79, K13} have indicated that the annihilation distance for edge dislocations is around 1.5 nm during stage I and stage II deformation. In the case of screw dislocations, the annihilation distances were much larger due to cross-slip: in the range of 10 - 15 nm during stage I and below 50 nm in stage III. Thus, an average value of $y_c$ = 15 nm was selected.

The dislocation MFP can be expressed as \cite{AS78, KDH08}

\begin{equation}\label{MFPd}
\ell^{\alpha}=\frac{K}{\sqrt{\displaystyle\sum_{\beta\neq\alpha}^{}{\rho^{\beta}}}},
\end{equation}

\noindent where $\rho^{\beta}$ is the total dislocation density on a latent system $\beta$ and $K$  is a dimensionless constant. In the case of Cu,  $K$ = 6 was obtained from the experimental relationship between the dislocation MFP and critical resolved shear stress for dislocation slip assuming that the later follows the Taylor model \cite{AS78, L06}. 

Experimental results \cite{NT81} as well as dislocation dynamics simulations \cite{LDH05, LDH07} have shown that the storage rate of dislocations increases as as the grain size decreases and this behavior can be explained following simple arguments \cite{KM03, LDH05}:  a dislocation loop that sweeps a cubic grain of dimensions $d \times d \times d$ leads to a shear strain $\Delta \gamma \approx b/d $. The associated increase in dislocation density  is given by $ \Delta \rho \approx 1 /d^2$  and thus $ \Delta \rho / \Delta \gamma \propto  1/bd$. Thus, the dislocation storage rate is not only governed by the dislocation MFP in the bulk but also by the grain size \cite{SK92, KM03, LDH05}. Moreover, dislocation dynamics simulations in polycrystals  with different grain size \cite{SDK10} have shown that the dislocation density is not constant within the grain but  increases as the distance to the grain boundary decreases. Based on these observations,  Lefebvre \cite{L06} modified eq. (\ref{evolution-dislocation}) to include the distance from the material point considered to the grain boundary, $d_b$, according to 

\begin{equation}\label{evolution-dislocation2}
\dot{\rho}^{\alpha}=\frac{1}{b} \left(\max \left(\frac{1}{\ell^{\alpha}},\frac{K_{s}}{d_b}\right)- 2y_c \rho^{\alpha} \right)| \dot{\gamma}^{\alpha}| ,
\end{equation} 

\noindent where $K_{s}$ is another dimensionless constant that controls the storage of dislocations on the grain boundary. Dislocation dynamics simulation of FCC crystals with different sizes have shown that $K_s$ $\approx$ 5 \cite{SDK10}. Thus, this physically-based, phenomenological modification of the Kocks-Mecking law can take into account the increase in dislocation density near the grain boundaries, which naturally leads to a grain size effect.

The strain hardening rate for the slip system $\alpha$, $\dot{\tau}_c^{\alpha}$,  can be obtained by differentiation of eq. (\ref{tauc}) with respect to time. Taking into account eqs. (\ref{MFPd}) and (\ref{evolution-dislocation2}),  this leads to

\begin{equation}
\dot{\tau}_c^{\alpha}=\sum_{\beta}h^{\alpha \beta}| \dot{\gamma}^{\beta} |
\end{equation}

\noindent where the hardening matrix $h^{\alpha \beta}$ is expressed as

\begin{equation}
h^{\alpha \beta}=\frac{\mu}{2}a^{\alpha \beta}\left(\sum_q a^{\alpha q} \rho^q\right)^{-\frac{1}{2}}\left \lbrace \max \left(\frac{1}{\ell^{\beta}},\frac{K_{s}}{d_b}\right)- 2y_c \rho^{\beta} \right \rbrace
\end{equation}

This constitutive model was implemented in Abaqus/Standard as a UMAT following the strategy presented in \cite{SL13}.

\section{Polycrystal homogenization framework}\label{sec2}

The mechanical behavior of the polycrystal is obtained by means of the finite element simulation of the deformation of a  Representative Volume Element (RVE) of the microstructure, following the standard procedures in computational homogenization \cite{MSS93, LLP04, LRS11, SLL14}. The cubic RVE is made up of a regular mesh of $N \times N \times N$ cubic finite elements or voxels (C3D8 elements in Abaqus with 8 nodes at the cube corners and full integration).

\begin{figure}
 \centering
 \includegraphics[scale=0.3]{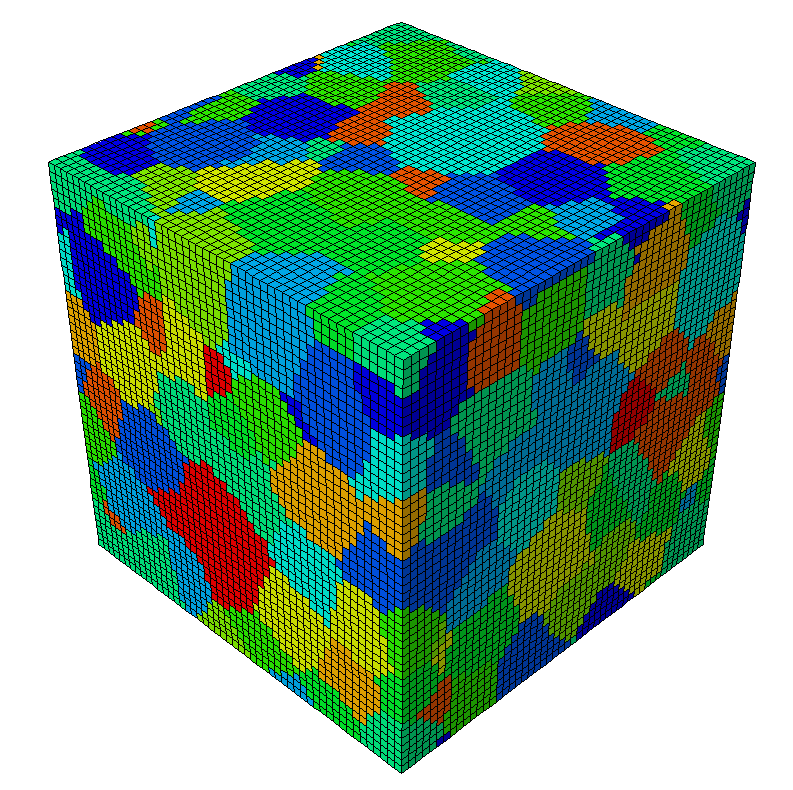}
\caption{Representative volume element of polycrystalline Cu containing 200 crystals discretized with 125000 cubic finite elements.} \label{RVE}
\end{figure}

The grain size distribution of the polycrystal followed a lognormal distribution  characterized by the average grain size, $d_g$, and the corresponding standard deviation, $d_{SD}$. The grains were equiaxed and the microstructure in the RVE was generated using Dream3D \cite{DREAM3D} (Fig. \ref{RVE}). Most simulations were carried out in RVEs with random texture but one set of analysis was carried out with the typical rolling texture of Cu to assess the influence of this factor on the Hall-Petch effect.

The microstructure of the RVE was periodic along the 3 directions of the RVE and periodic boundary conditions were applied to the cube faces according to 

\begin{equation} \label{BCs}
\begin{split}
\mathbf{u}(0,x_2,x_3)-\mathbf{u}(L,x_2,x_3) & = (\overline{\mathbf{F}}-\mathbf{1})L\mathbf{e}_1\\
\mathbf{u}(x_1,0,x_3)-\mathbf{u}(x_1,L,x_3) & = (\overline{\mathbf{F}}-\mathbf{1})L\mathbf{e}_2\\
\mathbf{u}(x_1,x_2,0)-\mathbf{u}(x_1,x_2,L) & = (\overline{\mathbf{F}}-\mathbf{1})L\mathbf{e}_3,
\end{split}
\end{equation}

\noindent where $L$ is the length of the cube size, $\mathbf{u}$ the displacement vector,  $\overline{\mathbf{F}}$ the far field macroscopic deformation gradient and $\mathbf{e}_i$, $i=1,2,3$ is the orthogonal basis with corresponding coordinates $x_i$, $i=1,2,3$. 

The far-field deformation gradient $\bar{\mathbf{F}}$ applied to the RVE is obtained by prescribing the displacements of three master nodes $M_i$ corresponding to three different faces of the RVE,

\begin{equation}
\mathbf{u}(M_i)=(\bar{\mathbf{F}}-\mathbf{I})\mathbf{e}_i.
\end{equation}

\noindent  If some components of the far-field deformation gradient are not known {\it a priori} (mixed boundary conditions, as in under uniaxial tension), the corresponding components of the effective stresses $\bar{\boldsymbol{\sigma}}$ are set instead. This is carried out by applying a nodal force $P_j$ to the master node $M_i$ and degree of freedom $j$ according to

\begin{equation}
P_j(M_i)=(\bar{\boldsymbol{\sigma}}\mathbf{e}_i)_j A_i
\end{equation} 

\noindent where $A_i$ is the projection of the current area of the face perpendicular to $\mathbf{e}_i$ in this direction. 

Finally, the macroscopic Cauchy stresses acting on any cube surface can be computed by dividing the reaction forces $F_j$ of the master nodes $M_i$ by the actual area of the face perpendicular to that master node $A_i$.

\begin{equation}
\bar{\sigma}_{ij}=\frac{F_j}{A_i}
\end{equation}

The constitutive equation developed in the previous section includes the distance  to the nearest grain boundary for each slip system.  This information was computed and stored at the beginning of the simulations for each slip system in each Gauss point. The deformation gradient in these simulations was small and it was assumed that this distance to the nearest grain boundary did not change during the analysis. 

The finite element simulations of the RVE to compute the polycrystal behavior were carried out in Abaqus/Standard \citep{A17} within the framework of the finite deformations theory with the initial unstressed state as reference.  The non-linear constitutive equation was integrated using a Newton-Raphson algorithm.

\section{Results and discussion}\label{sec3}

The computational homogenization strategy was used to analyze the influence of grain size on the tensile response of Cu polycrystals with average grains sizes in the range 10 to 80 $\mu$m. The elastic constants, strain rate sensitivity and reference strain rate of single crystal Cu are well known from the literature and shown in Table \ref{tab1}. The  parameters that control the hardening, storage and annihilation of dislocations during deformation were also determined for Cu using results in the literature from dislocation dynamics simulations and experiments and they are included in Table \ref{tab1}. All the simulations presented below were carried out at a constant strain rate of 7.0  10$^{-4}$ s$^{-1}$.

\begin{table}[ tp ]%
\caption {Parameters of the dislocation-based crystal plasticity model for Cu single crystals}\label{tab1}
\begin{center}
   \begin{tabular}{ l  l  l  l l}
   \hline
 {\it Elastic constants}  \cite{L06}: & $C_{11}$= 168.4  GPa & $ C_{12}$ = 121.4 GPa  & \\
 & $C_{44}$= 75.4  GPa & \\ 
     Shear modulus \cite{KH15}: & $\mu$ = 30.5 GPa & & & \\
     \hline
      
      {\it Viscoplastic parameters} \cite{EBG04}: & & \\ 
    
     reference shear strain rate  & $\dot{\gamma}_0$ =2.3  10$^{-4}$ s$^{-1}$ & & \\ 
     Strain rate sensitivity coefficient & $m$ = 0.05 & &\\
     \hline
     {\it Dislocation parameters} \cite{K13}:\\
      Burgers vector  &$b=2.56 \, 10^{-10}$ m  & & \\
     Annihilation distance  & $y_c$ = 15 nm  &  &\\
      \hline
      {\it Interaction coefficients:} & & & \\
      Self interaction \cite{DHK08}&$0.122 $ &\\
      Coplanar interaction \cite{DHK08}&$0.122 $ &\\
      Collinear interaction \cite{BCB13} &$0.657$&\\      
      Glissile junction \cite{DHK08}&$0.137$& \\
      Hirth lock \cite{BCB13} 	& $0.084$&\\
      Lomer-Cottrell  lock \cite{BCB13}	& $0.118$	&\\
    \hline
     Dislocation storage  \cite{AS78} & $K$ = 6 & \\
     Grain boundary storage  \cite{SDK10} & $K_s$ = 5 & \\
     \hline
      \hline
   \end{tabular}
 \end{center}
 \end{table}

\begin{figure}
 \includegraphics[scale=0.8]{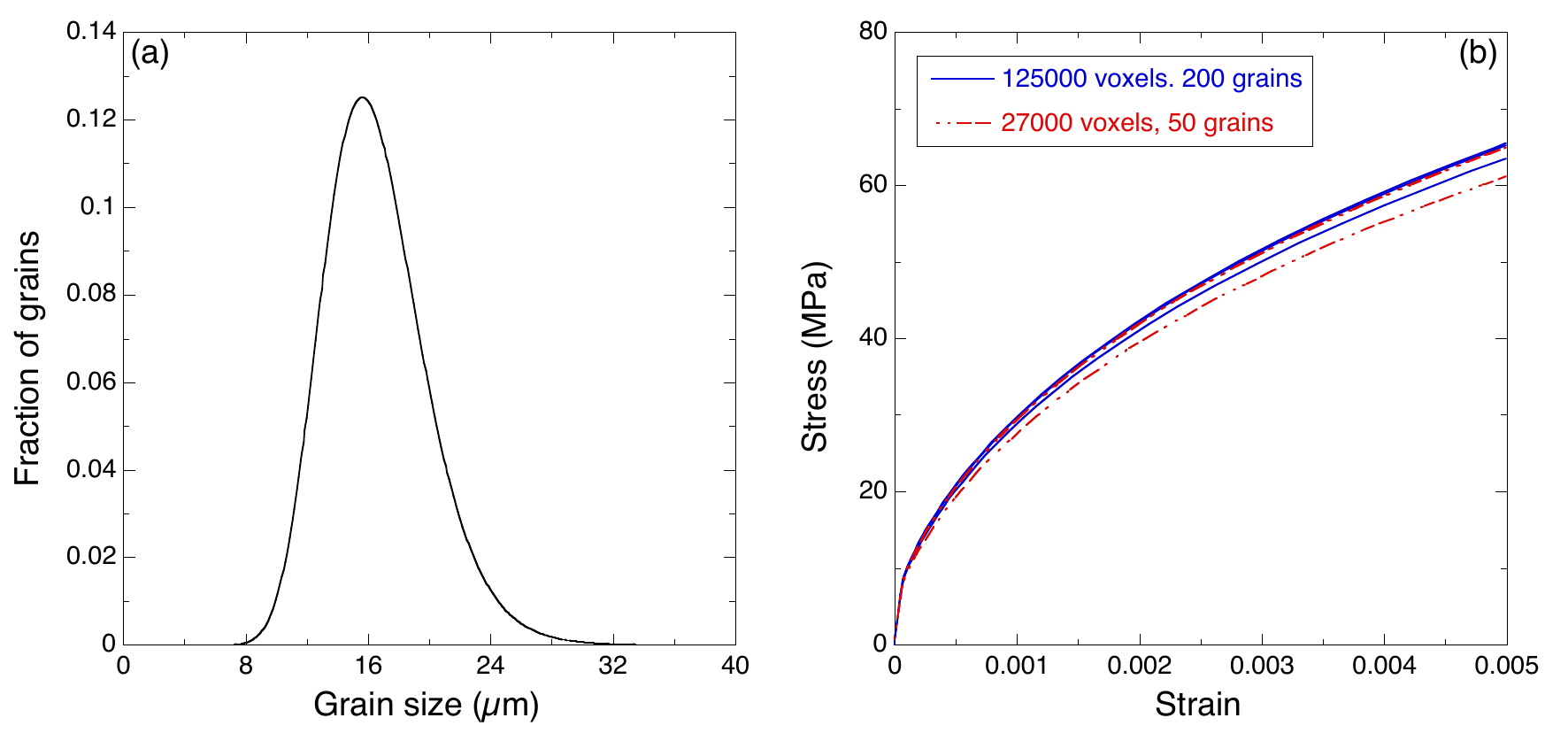}
 \caption{(a) Grain size distribution corresponding to an average grain size, $d_g$ = 20 $\mu$m, and a standard deviation $d_{SD}$ = 4 $\mu$m. (b) Influence of the discretization and number of grains in the RVE  on the stress-strain curve of polycrystalline Cu. The results of three different realizations of the grain size distribution are shown for each discretization.}
 \label{RVEsize}
\end{figure}

In order to check the critical size of the RVE, preliminary simulations were carried out using 27000 ($N$ = 30) voxels and 50 grains and 125000 ($N$ = 50) voxels and 200 grains in the RVE. These numbers were selected so the same number of voxels was used to discretize each grain in both models. The initial dislocation density in each slip system was  10$^{11}$ m$^{-2}$, leading to a total initial dislocation density $\rho_i$ = 1.2 10$^{12}$ m$^{-2}$ and the grain size distribution ($d_g$ = 20 $\mu$m, $d_{SD}$ = 4 $\mu$m) is depicted in Fig. \ref{RVEsize}a). Three different grain size realizations with random texture were simulated for each discretization and the corresponding stress-strain curves are plotted in Fig. \ref{RVEsize}b). The differences in the stress-strain curves among the three realizations for each discretization are small (below 5\% in the case of the finest discretization) as well as the differences in the curves obtained with 27000 and 125000 voxels. These results indicate that homogenized properties are independent of the RVE size and  can be  used to obtain the effective properties of the polycrystals, in agreement with previous results \cite{EFG07, SL13, HLD14}.

All the stress-strain curves reported below were obtained with RVEs including 125000 voxels and 200 grains. Each grain in the polycrystal was discretized with $\approx$ 625 voxels and the voxel  length was $\approx$ 1 $\mu$m in the case of a polycrystal with an average grain size of 10 $\mu$m, which is equivalent to the average distance between dislocations (1/$\sqrt{\rho}$) for a dislocation density of 10$^{12}$ m$^{-2}$. The finite element model assumes that the plastic deformation is homogeneously distributed in all the voxels within the grain but this assumption may not represent adequately the inhomogeneous plastic deformation that occurs in small grains (below 10 $\mu$m) with low dislocation densities. Moreover, the standard Taylor model (see eq. (\ref{tauc})) is no longer valid below this grain size for dislocation densities $<$ 10$^{12}$ m$^{-2}$, according to the dislocation dynamics simulations \cite{E15}. Thus, the minimum average grain size of the polycrystals in the simulations was 10 $\mu$m and the minimum value of the initial dislocation density 1.2 10$^{12}$ m$^{-2}$. 

\subsection{Influence of the grain size on the flow stress of Cu polycrystals}

\begin{figure}
\centering
 \includegraphics[scale=0.8]{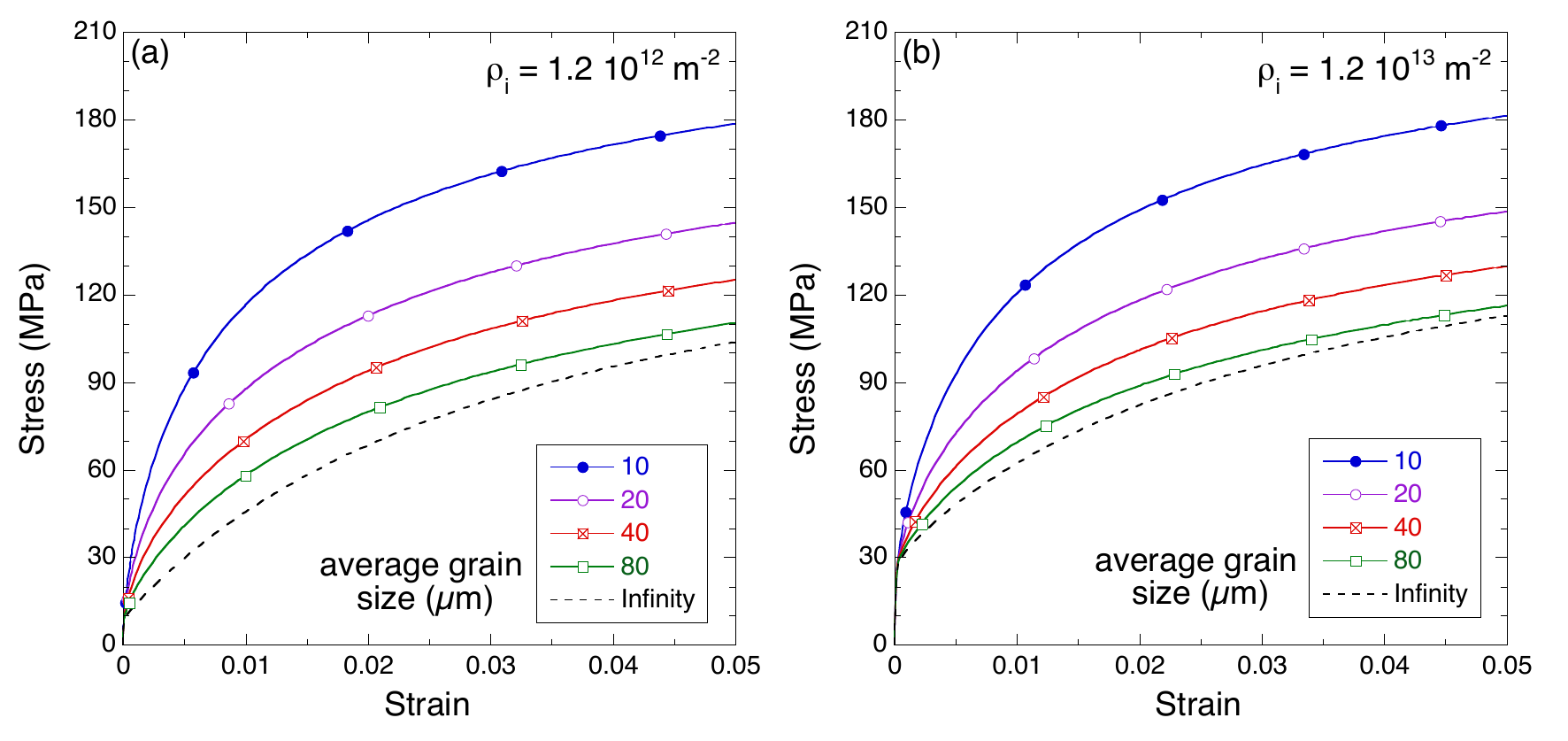}
  \includegraphics[scale=0.8]{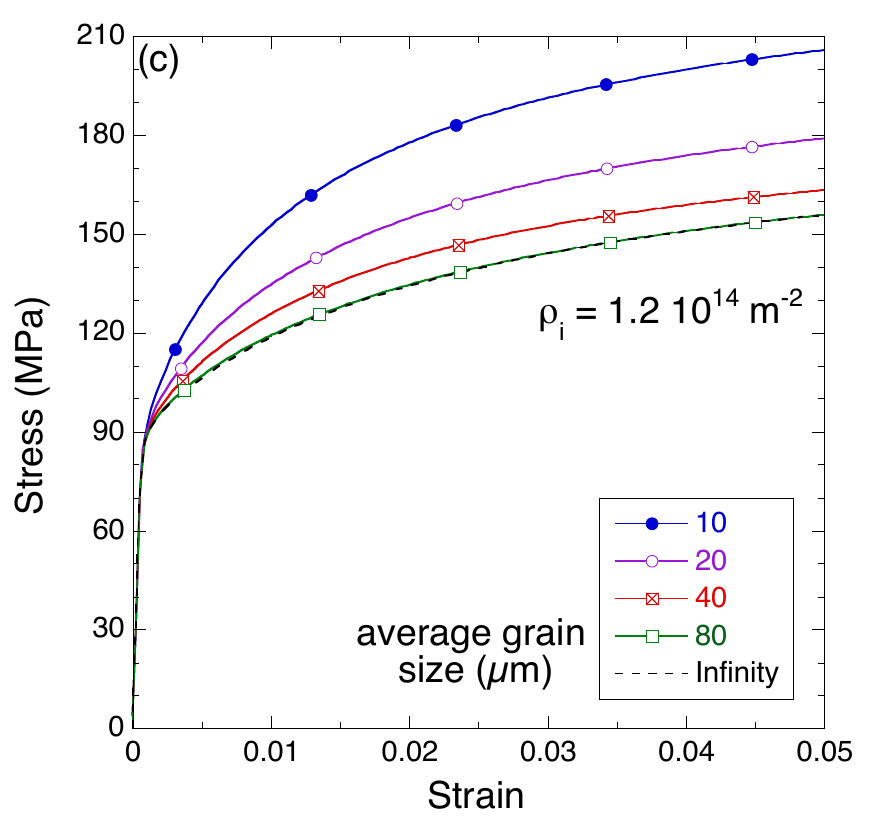}
 \caption{Stress-strain curves of Cu polycrystals as a function of the average grain size. (a) Initial dislocation density, $\rho_i$ = 1.2 10$^{12}$ m$^{-2}$. (b) Initial dislocation density, $\rho_i$ = 1.2 10$^{13}$ m$^{-2}$. (b) Initial dislocation density, $\rho_i$ = 1.2 10$^{14}$ m${-2}$. The broken lines stand for the results obtained when dislocation storage at the grain boundaries is not included in the model.}
 \label{fig:SE}
\end{figure}

The tensile behavior of polycrystals with $d_g$ = 10, 20, 40 and 80 $\mu$m and $d_{SD}$ = 0.2 $d_g$ was computed for three initial values of the dislocation density, $\rho_i$ = 1.2 10$^{12}$ m$^{-2}$, 1.2 10$^{13}$ m$^{-2}$ and 1.2 10$^{14}$ m$^{-2}$, and the corresponding stress-strain curves are plotted in Figs. \ref{fig:SE}a), b) and c), respectively. The results obtained neglecting the effect of dislocation storage at the grain boundaries ($K_s$ = 0) are also plotted as broken lines in these figures. The stress-strain curves in this case were superposed, regardless of the grain size, because the constitutive equation does not  include any size-dependent term. Thus, they were considered representative of a polycrystal with "infinite" grain size.

The initial flow stress of the polycrystals in Fig. \ref{fig:SE} is independent of the grain size and depends only on the initial dislocation density. However, the initial strain hardening rate after yielding increases rapidly as the grain size decreases, following the experimental trends,  due to the accumulation of dislocations at the grain boundaries. The strengthening induced by grain boundaries is associated to the region near the grain boundary in which the storage of dislocations induced by the presence of the boundary reduces the actual dislocation MFP. The thickness of this region and the magnitude of the size effect mainly depends on $K_s$, which controls the storage of dislocations at the grain boundaries. Thus, it is obvious from these simulations that the grain size as well as the initial dislocation density are key parameters to take into account the influence of grain boundaries on the strengthening of polycrystals. 

The strain hardening rate drops very rapidly for applied strains  $>$ 2\%, and this reduction is faster in the polycrystals with small grain size. This phenomenon is controlled by the annihilation of dislocations in the grain boundaries and depends on the  critical distance for dislocation annihiliation, $y_c$. Finally, the hardening rate seems to be independent of the grain size for applied strains $>$ 4\% (and very similar to that found in polycrystals with infinite grain size), indicating that the storage and annihilation of dislocations at the grain boundaries have reached an steady-state which is independent of the grain size at this stage. 

\begin{figure}
\includegraphics[scale=0.24]{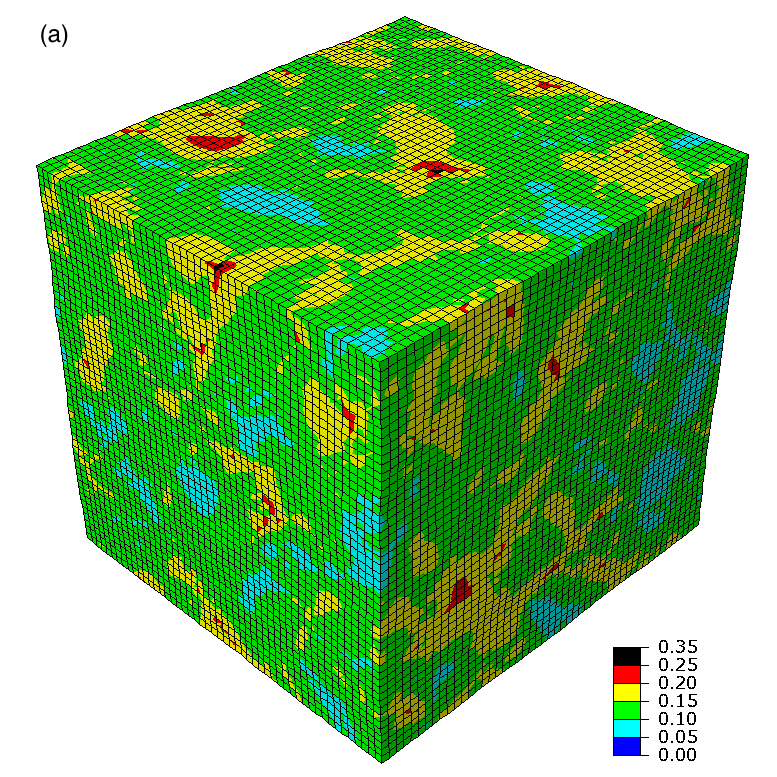}
 \includegraphics[scale=0.24]{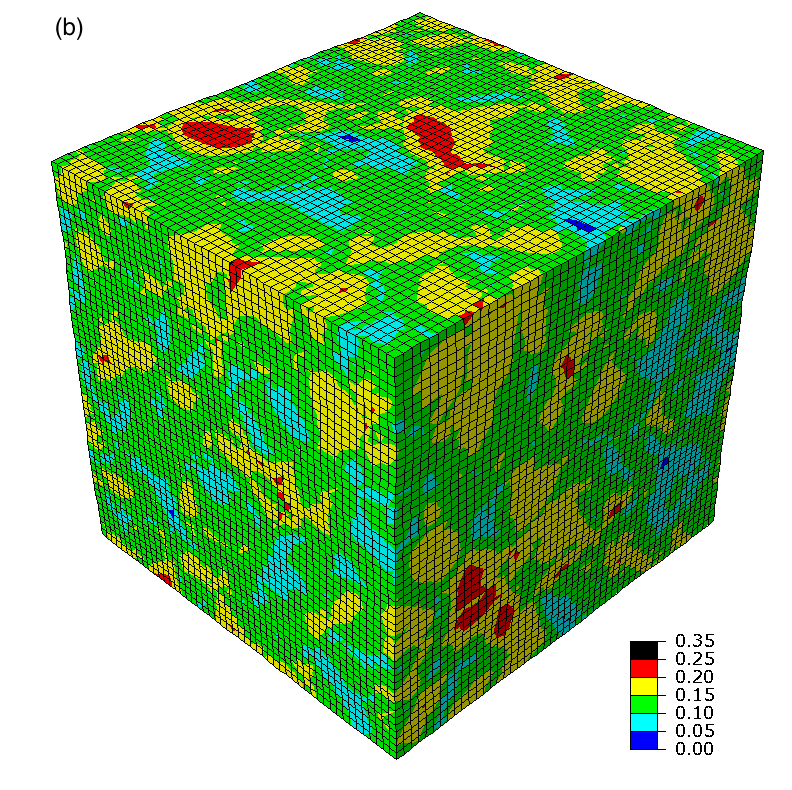}
 \centering
  \includegraphics[scale=0.24]{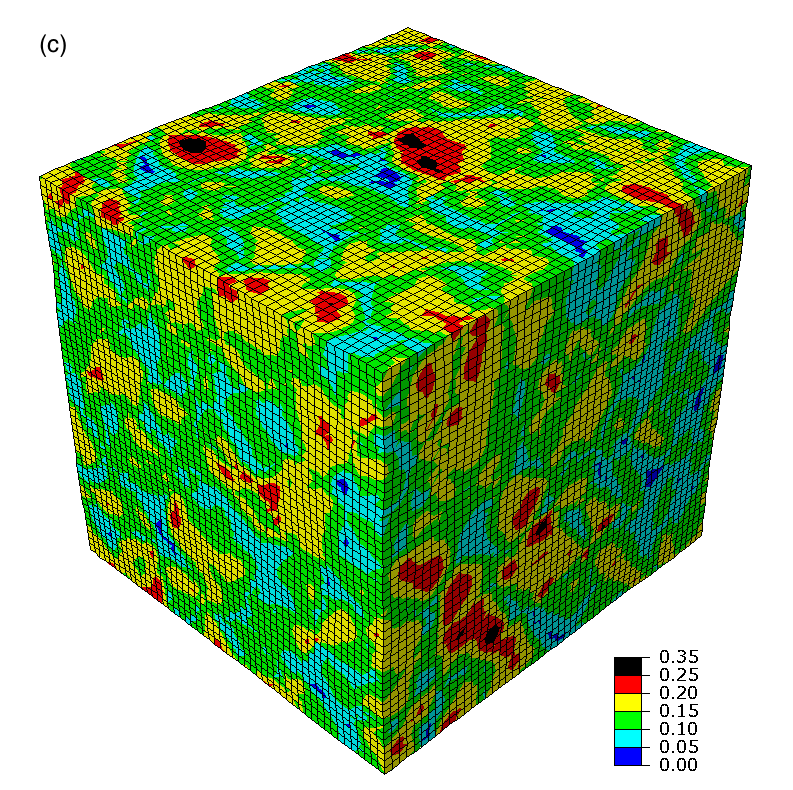}
 \caption{Contour plot of the accumulated plastic slip on all the slip systems, $\Gamma$, for polycrystals with different grain size and an initial dislocation density of 1.2 10$^{12}$ m$^{-2}$. (a) "Infinite" grain size. (b) $d_g$ = 40 $\mu$m. (c)  $d_g$ = 10 $\mu$m. The far-field applied strain was  5\% in all cases.}
 \label{fig:PE}
\end{figure}

\begin{figure}
\includegraphics[scale=0.24]{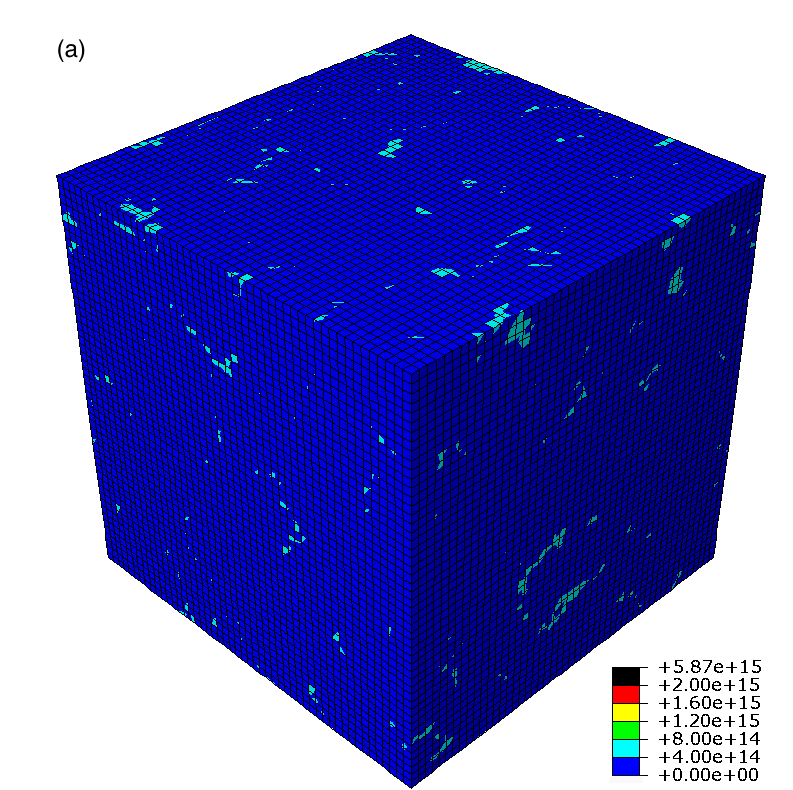}
 \includegraphics[scale=0.24]{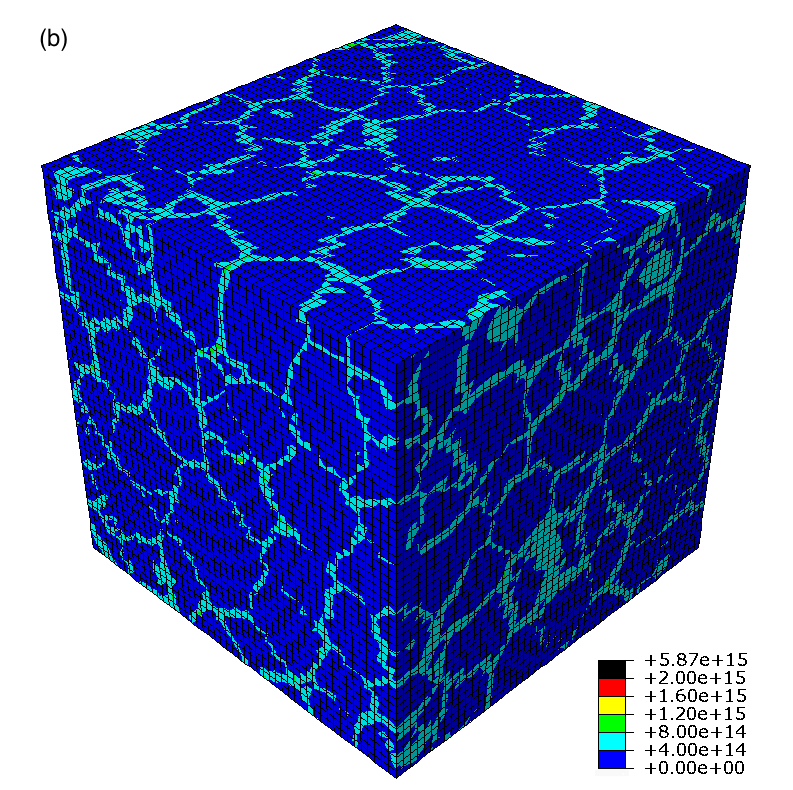}
 \centering
  \includegraphics[scale=0.24]{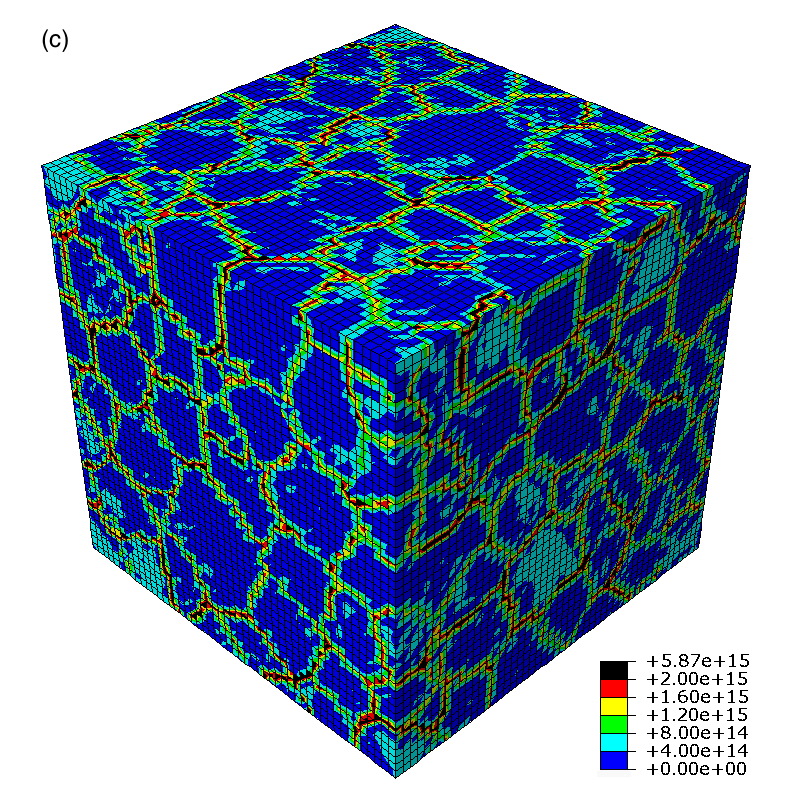}
 \caption{Contour plot of the total dislocation density in all the slip systems for polycrystals with different grain size and an initial dislocation density of 1.2 10$^{12}$ m$^{-2}$. (a) "Infinite" grain size. (b) $d_g$ = 40 $\mu$m. (c)  $d_g$ = 10 $\mu$m. The far-field applied strain was  5\% in all cases.}
 \label{fig:DD}
\end{figure}

\begin{figure}
\includegraphics[scale=0.28]{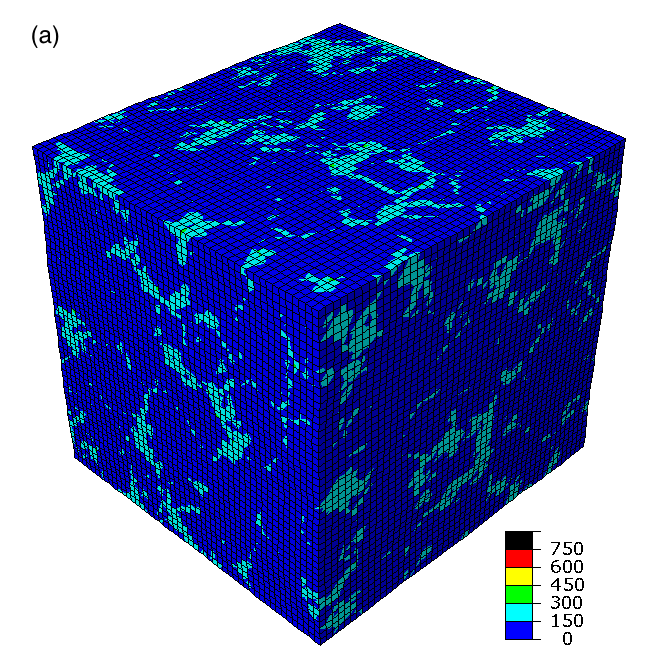}
 \includegraphics[scale=0.24]{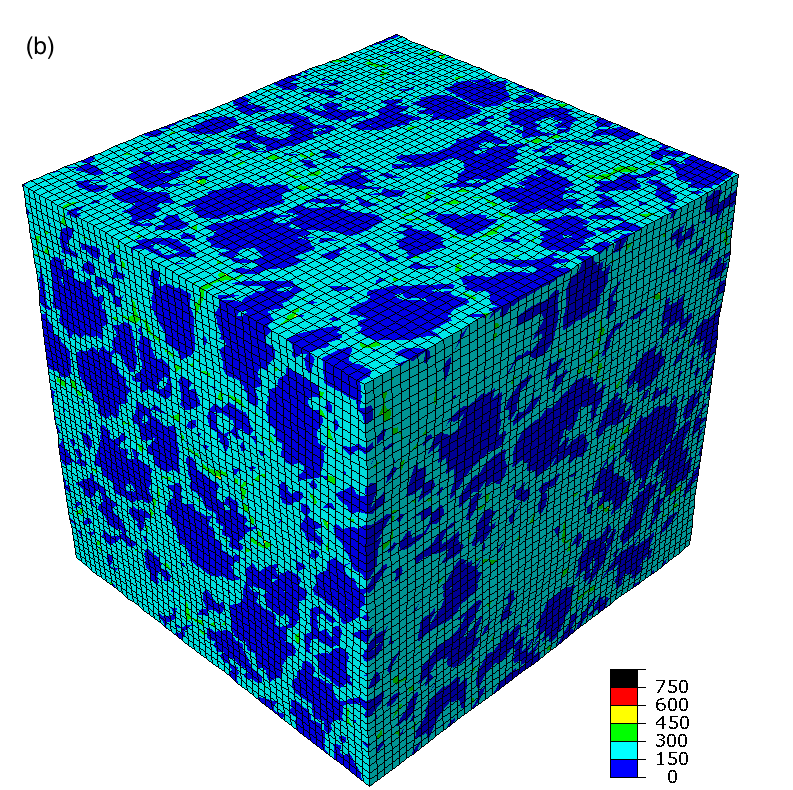}
 \centering
  \includegraphics[scale=0.24]{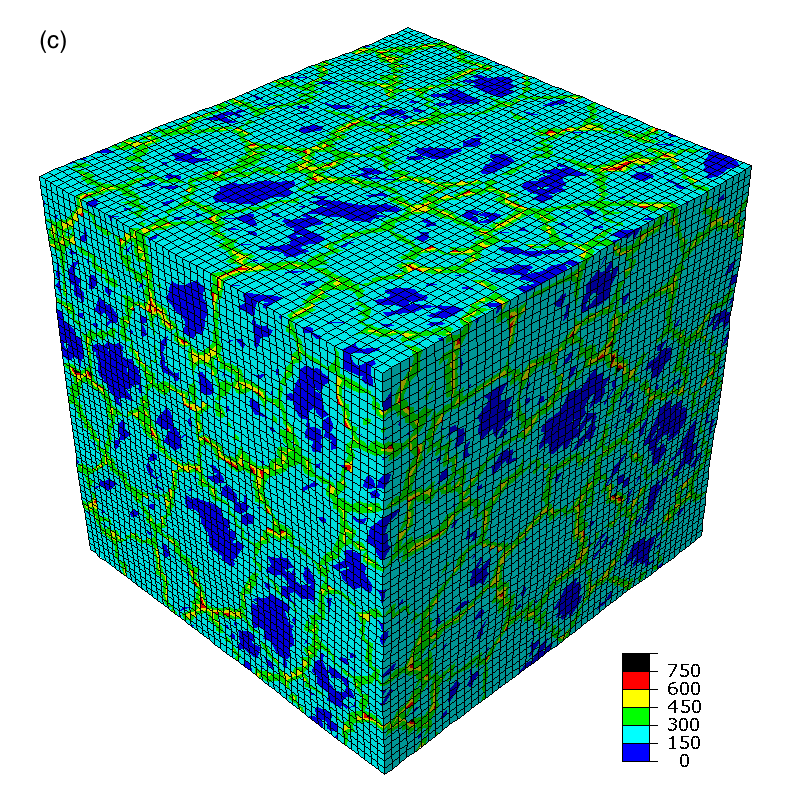}
 \caption{Contour plot of the Von Mises stress for polycrystals with different grain size and an initial dislocation density of 1.2 10$^{12}$ m$^{-2}$. (a) "Infinite" grain size. (b) $d_g$ = 40 $\mu$m. (c)  $d_g$ = 10 $\mu$m. The far-field applied strain was  5\% in all cases. Stresses are expressed in MPa.}
  \label{fig:VM}
\end{figure}

The influence of the grain size on the deformation pattern of the polycrystal can be assessed from Figs. \ref{fig:PE}, \ref{fig:DD} and \ref{fig:VM}, in which the contour plot of the accumulated plastic slip on all the slip systems ($\Gamma = \sum_\alpha \int{ | \dot\gamma^\alpha | \rd t}$), the total dislocation density and  the Von Mises stress are plotted, respectively, for polycrystals with average grain sizes of 10 $\mu$m, 40 $\mu$m and "infinite" grain size. In the case of polycrystals with "infinite" grain size, the accumulated plastic slip, the dislocation density and the Von Mises stress are fairly homogeneous throughout the microstructure, Figs. \ref{fig:PE}a), Fig. \ref{fig:DD}a) and \ref{fig:VM}a). 
Isolated "hot spots" in which the dislocation density and the Von Mises stress are higher can be seen in a few grains boundaries as a result of the elastic anisotropy and of the incompatibility in the plastic deformation between grains with different orientation. Nevertheless, their contribution to the overall flow stress of the polycrystal is negligible. On the contrary,  the plastic strain distribution becomes more heterogeneous throughout the microstructure as the grain size decreases, Figs. \ref{fig:PE}b) and c). Thus, plastic deformation tends to localize in large grains which are suitable oriented for slip, while it remains low in small grains because of the constraint of the grain boundaries.
This is clearly shown in Fig. \ref{fig:DD}, in which the dislocation densities are plotted for the three cases. They are homogeneous and around 10$^{14}$ m$^{-2}$ in most of the microstructure in the simulations  with "infinite" grain size, Figs. \ref{fig:DD}a), and much higher around the grain boundaries in the other two cases, reaching values  $>$ 10$^{15}$ m$^{-2}$ when the average grain size is around 10 $\mu$m, Figs. \ref{fig:DD}c). As a result, the stresses necessary to promote plastic deformation at the grain boundaries increased with respect to the stresses within the grains and the contour plots of the Von Mises stresses show very clearly the network of grain boundaries in the polycrystal, Figs. \ref{fig:VM}b) and c). The volume of material affected by this strengthening mechanism (as well as the maximum stress values) increase as the average grain size decreases, leading to the grain size effect on the flow stress.

\subsection{Comparison with experiments}

\begin{figure}
 \includegraphics[scale=0.8]{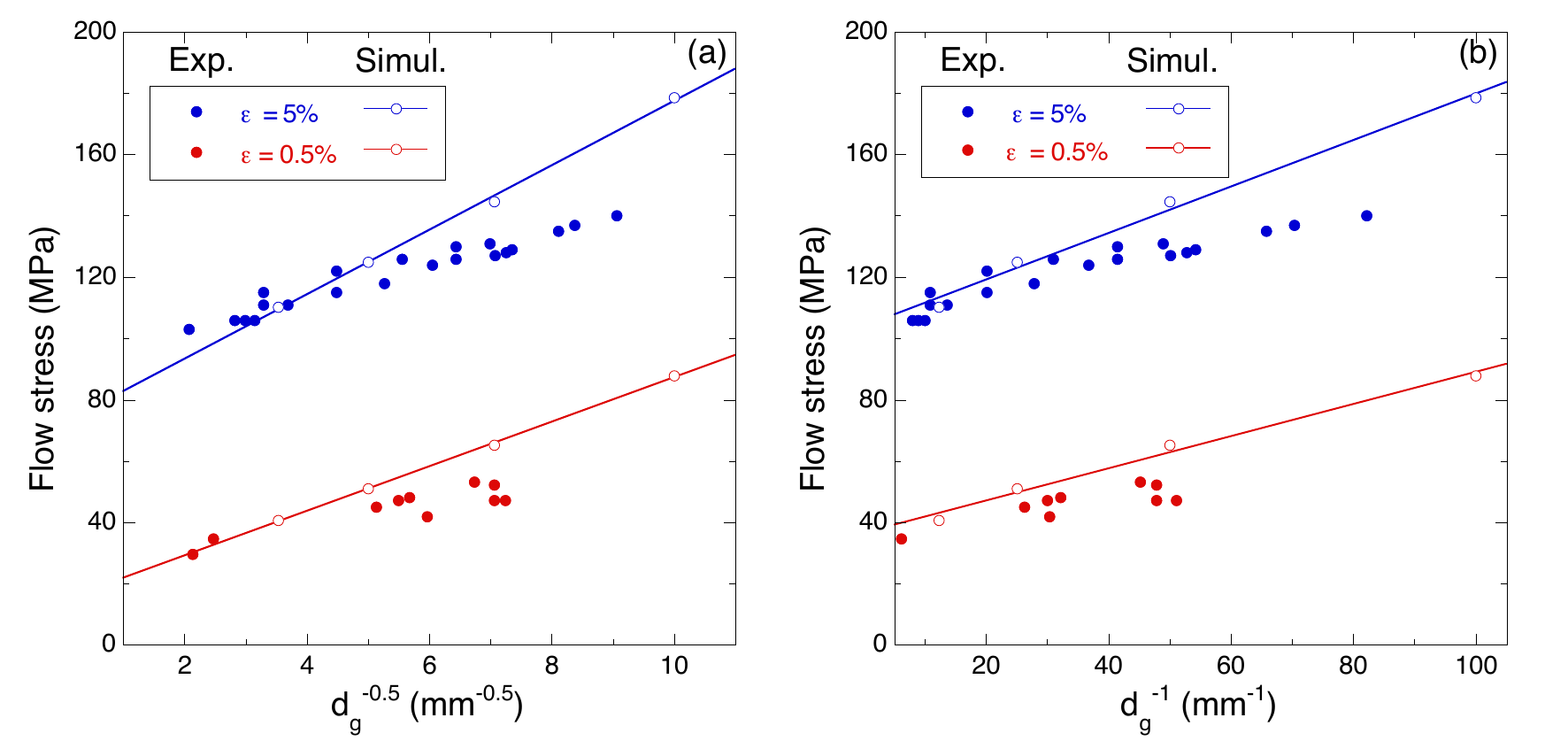}
 \caption{Influence of grain size on the flow stress of polycrystalline Cu after 0.5\% and 5¼\% tensile deformation. (a) Flow stress {\it vs.} $d_g^{-0.5}$. (b) Flow stress {\it vs.} $d_g^{-1}$. The experimental data  for an applied strain of 0.5 \% can be found in \cite{ACD62} while those corresponding to an applied strain of 5\% were obtained from \cite{HR82}.  The simulations were carried out using the parameters in Table 1 with an initial dislocation density of 1.2 10$^{12}$ m$^{-2}$.}
 \label{fig:HP}
\end{figure}

One critical test of the approach presented is its ability to provide a good estimation of the experimental evidence, taking into account that there are not adjustable parameters in the model. Li {\it et al.} \cite{LBD16} reviewed recently the experimental results available in the literature on the effect of grain size in the flow stress of polycrystalline Cu 
and those from Armstrong {\it et al.} \cite{ACD62} for an applied strain of 0.5\%  and from Hansen and Ralph \cite{HR82} for an applied strain of 5\% could be directly compared with the simulations in this paper. They are shown in Figs. \ref{fig:HP}a) and b) in which the flow stress after 0.5\% and 5\% applied strain is plotted as a function of $d_g^{-0.5}$ and $d_g^{-1}$, respectively. The results of the polycrystal homogenization simulations were carried out using the parameters in Table 1 and an initial dislocation density of 1.2 10$^{12}$ m$^{-2}$, which corresponds to a well-annealed polycrystal. 

It should be noted that the experimental data and the numerical predictions of the flow stress can be fitted to both $ d_g^{-0.5}$ and $d_g^{-1}$ within the range of grain sizes and applied strains studied. There is no information in the experimental report about the initial dislocation density but the Cu polycrystals were well annealed, so values of $\rho_i$  $\approx$ 10$^{12}$ m$^{-2}$ are reasonable. 
The numerical results obtained with this initial dislocation density are very close to the experimental data for grain sizes $\ge$ 20 $\mu$m although they overestimate slightly the flow stress at an applied strain of 0.5 \%. This latter difference may be explained by the fact that the grain boundary strengthening model  in the constitutive equation assumes that all grain boundaries store dislocations and does not take into account  the orientation of the crystals at both sides of the grain boundary. However, the contribution of some grain boundaries to the storage of dislocations is minimum because slip transfer between neighbour grains can be easily accommodated. The anisotropy of grain boundaries from the viewpoint of dislocation transmission and storage is very important for applications where the relative grain boundary fraction is significant, e.g. ultra fine-grained metals, thin films, micro-devices and in low symmetry crystals (because of the limited number of slip systems and the differences in the critical resolved shear stresses among the different systems) but it is very challenging from the simulation viewpoint \cite{BMK13}. However, the influence of this mechanism is more limited in FCC polycrystals and, thus, the model predictions for FCC Cu are in good agreement with the experimental data.

The model tends to overestimate the flow stress of the polycrystals with an average grain size of 10 $\mu$m and this difference can be attributed to two factors, Firstly, the overestimation of the strengthening effect of the grain boundaries by neglecting easy slip transfer, as indicated above. Secondly, the finite element crystal plasticity model may not represent adequately the inhomogeneous plastic deformation that occurs in small grains (below 10 $\mu$m) with low dislocation densities because the voxel size is equivalent to the average dislocation distance.

\subsection{Scaling laws for the flow stress}

\begin{figure}
\centering
 \includegraphics[scale=0.8]{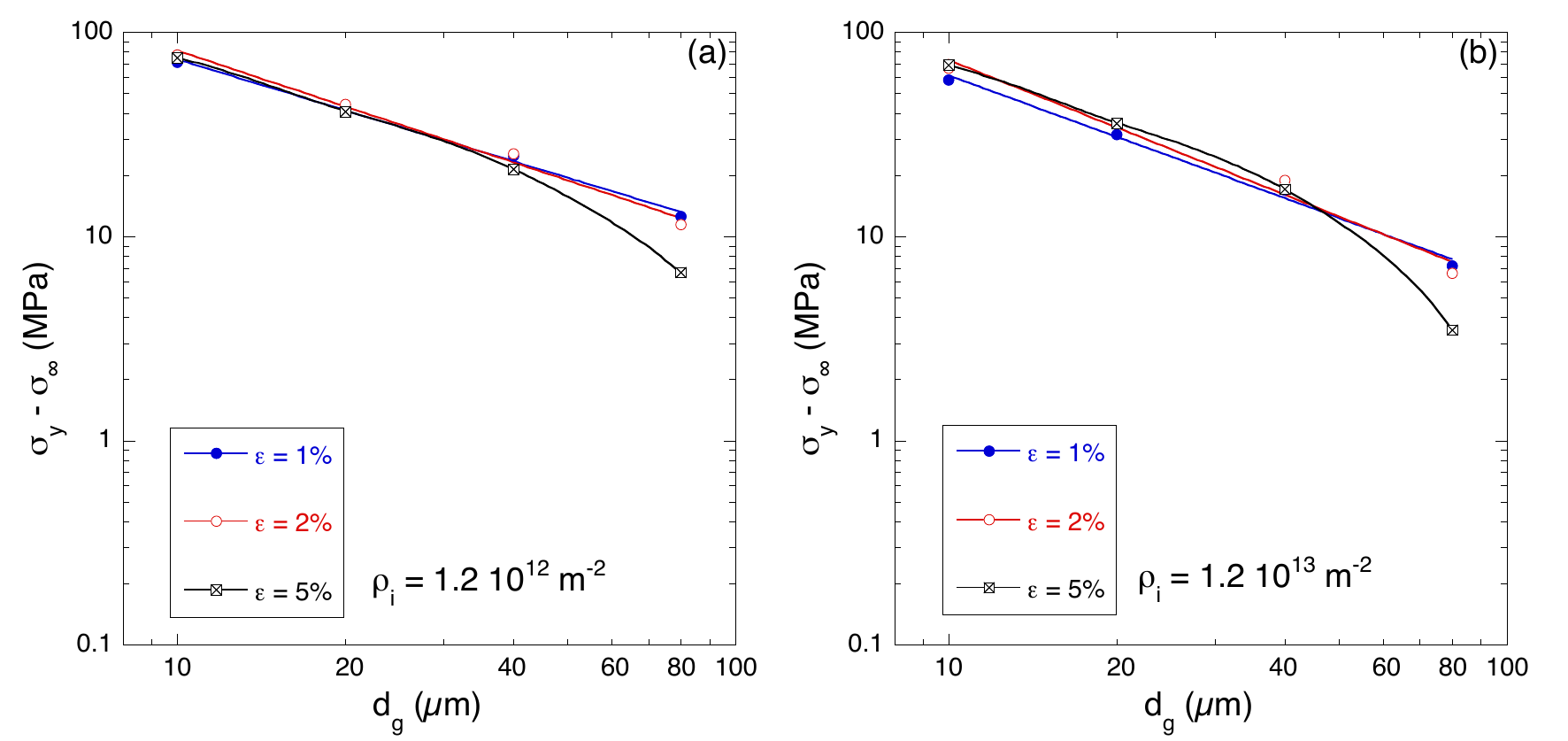}
  \includegraphics[scale=0.8]{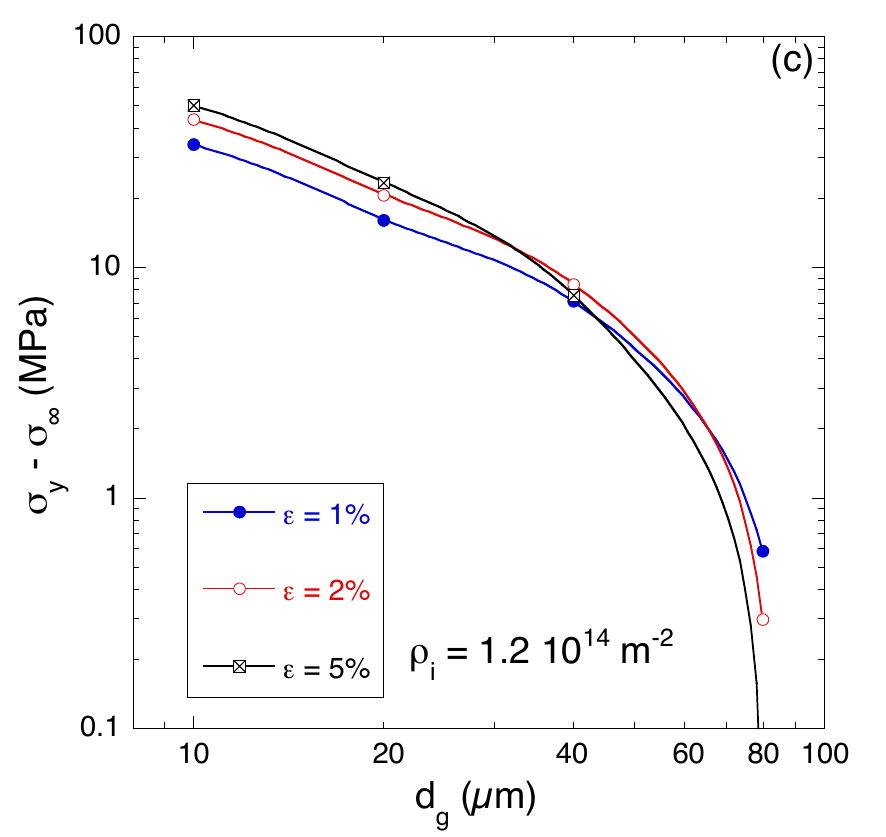}
 \caption{Strengthening provided by grain boundaries, $\sigma_y-\sigma_\infty$, as function of the grain size, $\bar d_g$, plotted in bilogarithmic coordinates for different values of the far-field applied strain $\epsilon$. (a) $\rho_i$ = 1.2 10$^{12}$ m$^{-2}$. (b) $\rho_i$ = 1.2 10$^{13}$ m$^{-2}$. (c) $\rho_i$ = 1.2 10$^{14}$ m$^{-2}$.}
 \label{fig:x}
\end{figure}

As indicated in the introduction, the experimental results for the effect of grain size on the flow strength of polycrystals are often approximated by  a generalized Hall-Petch equation, 

\begin{equation}
(\sigma_y-\sigma_\infty) = C d_g^{-x}
\label{eq:x}
\end{equation}

\noindent where $\sigma_y$ is the polycrystal flow stress at a given applied strain, $\sigma_\infty$  the flow stress of the polycrystal with "infinite" grain size at the same applied strain and $C$ and $x$ are materials constants  with $ 0 < x \le 1$ \cite{RP86}. It should be noted, however, that large discrepancies are found in the experimental literature in the value of $x$ even for nominally identical metals and alloys \citep{RP86} and the  simulations in this paper can provide valuable information about the range of validity of eq. (\ref{eq:x}). To this end, the results of the numerical simulations for $\sigma_y-\sigma_\infty$ {\it vs.} the average grain size, $d_g$, are plotted in bilogarithmic coordinates in Figs. \ref{fig:x}a), b) and c) for microstructures with initial dislocation densities of 1.2 10$^{12}$ m$^{-2}$, 1.2 10$^{13}$ m$^{-2}$ and 1.2 10$^{14}$ m$^{-2}$, respectively. The first value represents a well-annealed polycrystal with an initial yield stress of $\approx$ 10 MPa while the third one represents a work hardened material with an initial yield stress close to 100 MPa (Fig. \ref{fig:SE}). 
Data for three different values of the applied strain (1\%, 2.0\% and 5\%) are plotted in each figure. The numerical results for  $\rho_i$ = 1.2 10$^{12}$ m$^{-2}$  and   $\rho_i$ = 1.2 10$^{13}$ m$^{-2}$ (Figs. \ref{fig:x}a and b)  can be  well approximated by eq. (\ref{eq:x}) with $x$ $\approx$ 0.85 in the former and $x$ $\approx$ 1 in the latter for applied tensile strains of 1\% and 2\%. However, the linear relationship between log ($\sigma_y-\sigma_\infty$) and log ($d_g$) begins to disappear for both initial values of the dislocation density for $\epsilon$ = 5\%. The breakdown of the  the linearity expressed by eq. (\ref{eq:x}) in bilogarithmic coordinates is more obvious in the polycrystal with $\rho_i$ = 1.2 10$^{14}$ m$^{-2}$ (Fig. \ref{fig:x}c) and strengthening provided by the grain boundaries drops very rapidly for large grain sizes ($>$ 40 $\mu$m), regardless of the applied strain.

The results in Fig. \ref{fig:x} show the competition between the two mechanisms that dictate the effect of grain boundaries on the mechanical properties of the polycrystal. Strengthening is induced by the storage of dislocations at grain boundaries but this process is limited by the annihilation of dislocations around the grain boundaries when the dislocation densities reach very high values. The former process dominates when the initial dislocation density and the applied strain are small ($\rho_i \le$ 10$^{13}$ m$^{-2}$ and $\epsilon \le$ 2\%, respectively) and the strengthening provided by the grain boundaries follows the generalized Hall-Petch law expressed by eq. (\ref{fig:x}). However, annihilation of dislocations at the grain boundary becomes relevant for large applied strains ($\epsilon >$ 2\%) and/or high values of the initial dislocation density ($\rho_i >$ 10$^{14}$ m$^{-2}$) and the strengthening contribution of the the grain boundaries becomes irrelevant for large grain sizes ($>$ 40 $\mu$m), leading to a break down of the Hall-Petch effect. However, it should be noticed that it could have been possible to find a good correlation between the numerical results and eq. (\ref{eq:x}) if the data set was limited to grain sizes $\le$ 40 $\mu$m.

\begin{figure}
\centering
 \includegraphics[scale=0.8]{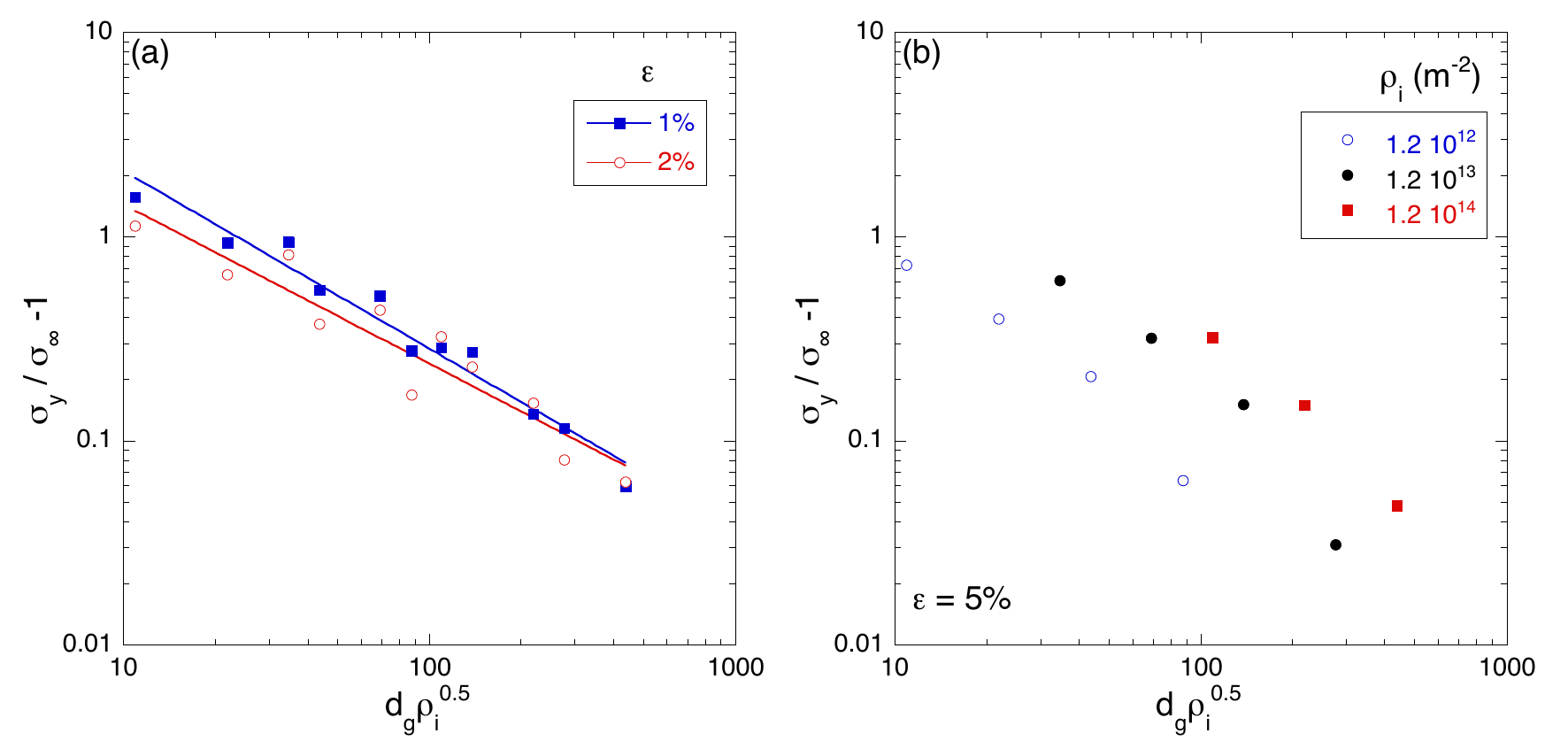}
 \caption{Strengthening provided by grain boundaries, $1- \sigma_y/\sigma_\infty$, as a function of the dimensionless parameter $d_g \sqrt{\rho_i}$ for polycrystals with different average grain size (10 $\mu$m $\le  d_g \le$ 80 $\mu$m)  and initial dislocation densities  (1.2 10$^{12}$ m$^{-2} \le \rho_i \le 1.2$ 10$^{14}$ m$^{-2}$).  (a) Applied strain $\epsilon$ = 1\% and 2\%. (b) Applied strain $\epsilon$ = 5\%. }
 \label{fig:SizeEffect}
\end{figure}

Thus, the simulations presented above indicate that the strengthening provided by grain boundaries in polycrystals do not depend only on the average grain size but also on the initial dislocation density. In the case of well annealed polycrystals (within initial dislocation densities  $<$ 10$^{13}$ m$^{-2}$),  the effect of grain size on the flow stress of FCC polycrystals can  be represented  by eq. (\ref{eq:x})  and the exponent $x$ is closer to 1 than to the original value of 0.5  proposed by Hall-Petch, in agreement with experimental observations \citep{DB13, DB14}. This scaling law breaks down, however, for FCC polycrystals with large initial dislocation densities ($>$ 10$^{14}$ m$^{-2}$) and grain sizes larger than 40 $\mu$m. This result is in agreement with theoretical results \cite{ZS14} and dislocation dynamics simulations \cite{E15} which show that the strengthening associated with size effects in plasticity, $\sigma_y-\sigma_\infty$ has to be expressed as

\begin{equation}
\sigma_y - \sigma_\infty = \sigma_\infty \Delta (d_g\sqrt{\rho})
\label{eq:SizeEffect}
\end{equation}

\noindent where $\Delta (d_g\sqrt{\rho})$ is a function of the ratio between two length scales: the physical length scale ($d_g$ in the case of polycrystals) and the average dislocation spacing ($1/\sqrt{\rho}$). This hypothesis is checked in Fig. \ref{fig:SizeEffect}, in which the strengthening of polycrystals due to the grain size, $1- \sigma_y/\sigma_\infty$, is plotted {\it vs.} $d_g \sqrt{\rho_i}$, where $\rho_i$ is the initial dislocation density. The simulation results for an applied strain of 1\% or 2\% are shown in Fig. \ref{fig:SizeEffect}a) and support this hypothesis. Regardless of the initial dislocation density, the strengthening  due to the grain size can be approximated by an expression on the form

\begin{equation}
\sigma_y/\sigma_\infty -1 =  C (d_g\sqrt{\rho_i})^{-x}
\label{eq:SizeEffect}
\end{equation}

\noindent where $C$ = 15.6 and $x$ = 0.87 for $\epsilon$ = 1\% and $C$ = 8.61 and $x$ = 0.78 for $\epsilon$ = 2\%. In the case of an applied strain of 5\% (Fig. \ref{fig:SizeEffect}b), the strengthening provided by the grain size decreases as $d_g\rho_i$ increases but the actual magnitude of $1- \sigma_y/\sigma_\infty$ also depends on the initial dislocation density. 

The results in Fig. \ref{fig:SizeEffect}a) point out that eq. \eqref{eq:SizeEffect} is able to capture the strengthening due to grain size for small applied strains when dislocation storage at the grain boundaries is the dominant mechanism and annihilation of dislocations at the grain boundaries was negligible. As the applied strain increases up to 5\%, dislocation annihilation at the grain boundaries starts to play an important role that is not included in the dimensionless parameter $d_g\rho_i$. Thus, the strength provided by grain boundaries still decreases as $d_g\rho_i$ increases at large applied strains but all the results do not collapse into a single line in bilogarithmic coordinates.

\subsection{Effect of microstructural features: grain size distribution and texture}

\begin{figure}
 \includegraphics[scale=0.8]{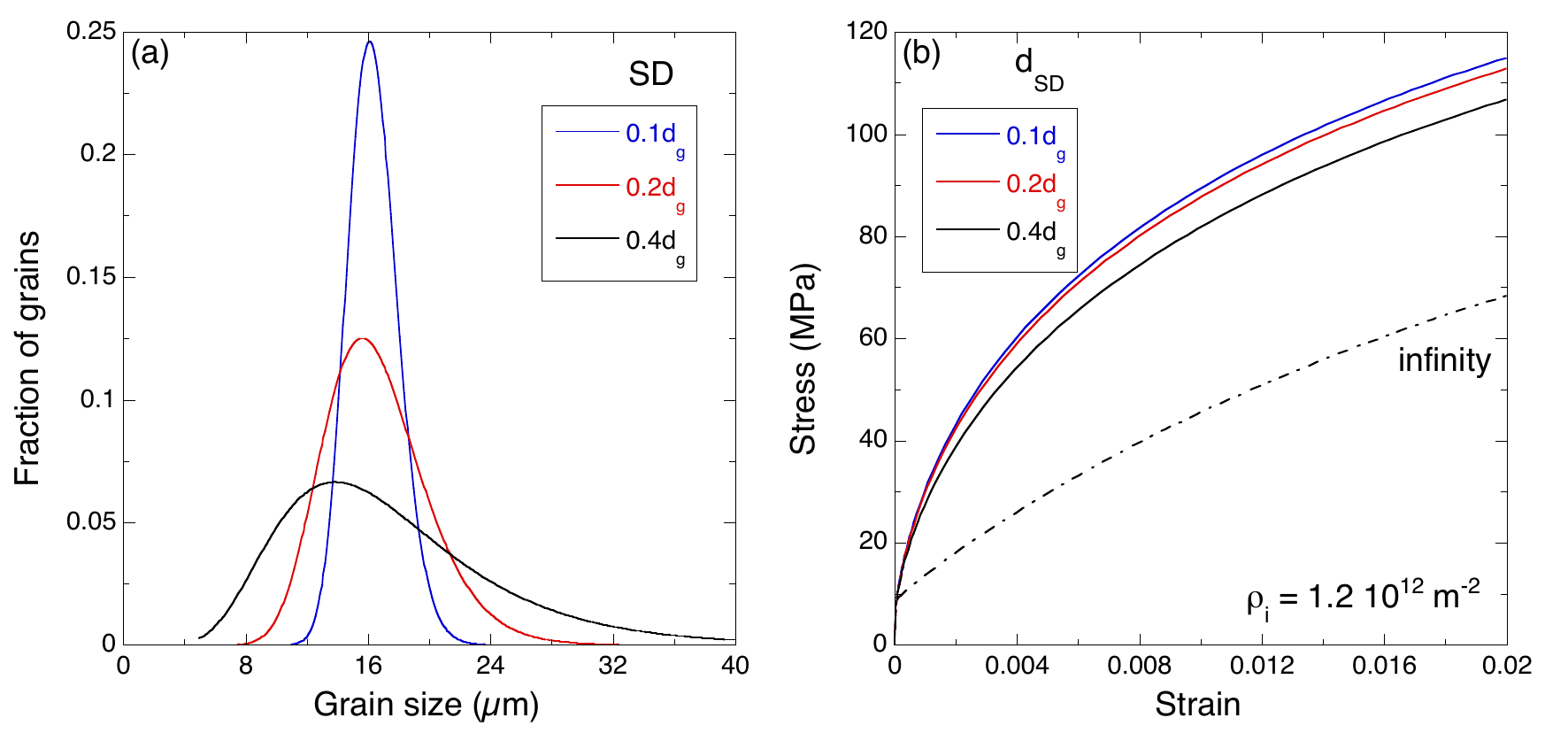}
 \caption{Grain size distribution for an average grain size, $d_g$ = 20 $\mu$m and three different values of the standard deviation of the grain size distribution, $d_{SD}= 0.1d_g$, 0.2$d_g$ and 0.4$d_g$. (b) Influence of the standard deviation of the grain size distribution on the flow stress of polycrystals. The broken line correspond to simulations in which dislocation storage at the grain boundaries was not included.}
 \label{fig:GSD}
\end{figure}

The polycrystal homogenization strategy allows the exploration of the influence of different microstructural factors on the strengthening due to the grain size and two of them (grain size distribution and texture) will be addressed in this section. RVEs with 200 grains and random texture were generated using three different grain size distributions indicated in Fig. \ref{fig:GSD}a). The average grain size, $d_g$, was constant and equal  to 20 $\mu$m in all cases but the standard deviation of the grain size distribution, $d_{SD}$, varied from 2 $\mu$m (a narrow distribution with $d_{SD}$ = 0.1$ d_g$) to 8 $\mu$m (a wide distribution with $d_{SD}$ = 0.4$d_g$). The influence of the width of the grain size distribution on the stress-strain curve is plotted in Fig. \ref{fig:GSD}b) for simulations carried out with an initial dislocation density of 1.2 10$^{12}$ m$^{-2}$. Two sets of simulations were carried out for each grain size distribution, with and without the effect of dislocation storage at the grain boundaries. The former are shown with solids lines and the latter with a broken line  because the grain size distribution did not influence  the flow stress of the polycrystal if the dislocation storage at the grain boundaries is not included in the model. However, narrower grain size distributions led to higher strengths if this effect was accounted for in the simulations. The effect of the width of the grain size distribution was not large but it was noticeable and this is another factor -- together with the initial dislocation density -- that may be responsible for the large scatter found in the experimental data of the grain size effect.

\begin{figure}
 \includegraphics[scale=0.3]{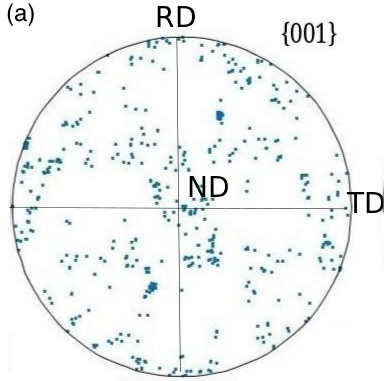}
  \includegraphics[scale=0.3]{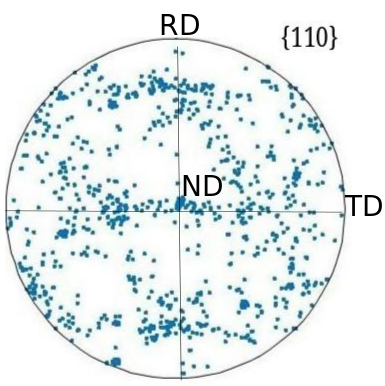}
 \includegraphics[scale=0.3]{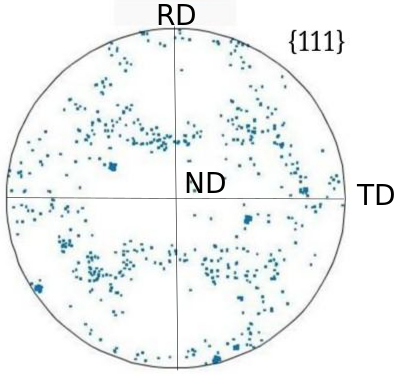}
\centering
  \includegraphics[scale=0.8]{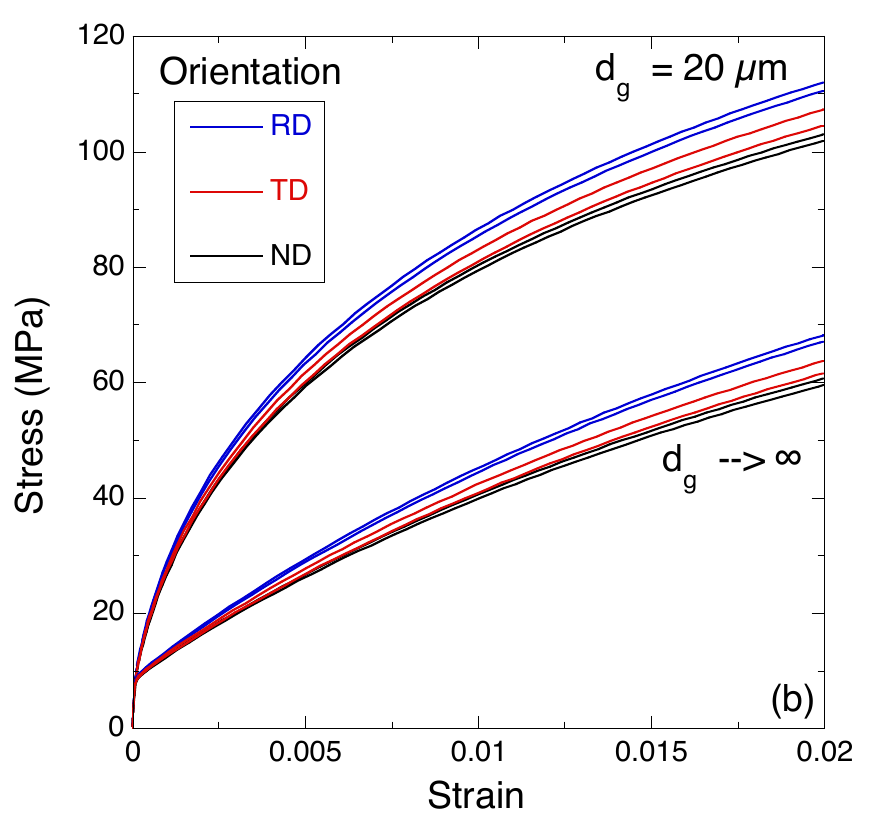}
 \caption{(a) Pole figures of rolled Cu. Each figure contains the orientation of the 200 grains in the RVE. (b) Influence of orientation on the stress-strain curve in rolled Cu. Two sets of curves (corresponding to $d_g$ = 20 $\mu$m and 'infinite' grain size) are presented for each orientation.}
 \label{fig:texture}
\end{figure}

The analysis of the influence of the initial texture on the grain size was carried out using an RVE with 200 grains.  Representative  \{001\},  \{110\} and \{111\} pole figures  are plotted in Fig. \ref{fig:texture}a) for the 200 grains in the RVE, which were obtained from the experimental texture of a rolled sample using a Monte Carlo lottery to assign the grain orientation within the RVE. They show the typical texture of Cu with respect to  RD, TD and ND (rolling, transverse and normal directions of the sheet), respectively. The \{111\} pole figure clearly indicates that the material is highly textured and that the \{111\} planes lie parallel to the rolling plane, which is a common rolling texture developed in pure FCC metals  \citep{BTS71, SSN02}.

The stress-strain curves obtained by computational homogenization along the rolling direction (RD), normal direction (ND) and transverse direction (TD) are plotted in Fig. \ref{fig:texture}b) for a grain size distribution characterized by  $\bar d_g$ = 20 $\mu$m and $d_{SD}$ = 4 $\mu$m and an initial dislocation density of 1.2 10$^{12}$ m$^{-2}$. The grains were assumed to be equiaxed (although it is known that this is not the case for rolled Cu) to account only for the grain orientation effect. Two simulations were carried out in each orientation with different texture realizations obtained by means of the Monte Carlo lottery. The corresponding stress-strain curves were very close in all cases, indicating that simulations with 200 grains were large enough to capture the effect of texture. In addition, polycrystal simulations in which the storage of dislocations at the grain boundaries was not accounted for are also included in this figure for the three orientations. The simulation results show that expected influence of the texture on mechanical behavior: the polycrystal was slightly stronger along the RD and the softest response was found along the ND. However, the differences in the flow stress are small, as is typical of FCC alloys because of the large number of slip systems, which lead to a rather isotropic plastic deformation even in the presence of a strong texture. Storage of dislocations at the grain boundaries led to a similar size effect in the three orientations and, thus, texture did not influence the magnitude of the grain boundary strengthening.

\section{Conclusions}

The influence of grain size on the mechanical response of FCC polycrystal has been studied  using a multiscale approach based on computational homogenization of the polycrystal behavior. The constitutive equation of the single crystals was given by a rate dependent physically-based crystal plasticity model in the context of finite strain plasticity. The critical resolved shear stress to produce plastic slip was obtained by a Taylor model in which the strengthening mechanisms due to dislocation/dislocation interactions and junctions were included. The generation and annihilation of dislocations in each slip system during deformation was given by the Kocks - Mecking model, which included an extra term to account for the dislocation storage at the grain boundaries. All the model parameters have a clear  physical meaning and could be obtained from dislocation dynamics simulations or experiments in the case of Cu.

The results of the numerical simulations showed that the yield stress was controlled by the initial dislocation density and was independent of the grain size. However, the strain hardening rate showed a strong effect of the average grain size, which was mainly attributed to the storage of dislocations at the grain boundaries. In the absence of this mechanism, the effect of the grain size on the mechanical behavior due to the elastic anisotropy and to the plastic deformation incompatibility between neighbour grains was negligible. The model predictions effectively captured the experimental trends for the grain size effect in polycrystalline Cu, validating the multiscale computational homogenization strategy. Two main factors were found to determine the strengthening provided by grain boundaries in polycrystals: the average grain size and the initial dislocation density. Other microstructural factors (width of the grain size distribution, texture) played a secondary role in the magnitude of the size effect. It was found that the scaling law $\sigma_y -\sigma_\infty \propto d_g^{-x}$ was fulfilled for well annealed polycrystals (with 0.85 $\le x \le$ 1) but did not hold in polycrystals with large initial dislocation densities ($>$ 10$^{14}$ m$^{-2}$) and grain sizes larger than 40 $\mu$m. These results explain the large differences in the literature in the proportionally constant and the exponent of the size effect law because very different values can be obtained as a function of the initial dislocation density or of the range of grain sizes explored. 
Finally, the simulation results showed that the contribution of the grain size to the strength followed a power-law  function of the dimensionless parameter $d_g\sqrt{\rho_i}$ for small values of the applied strain ($<$ 2 \%), in agreement with previous theoretical considerations for size effects in plasticity \cite{ZS14}.

\section{Acknowledgments}
This investigation was supported by the European Research Council under the European Union's Horizon 2020 research and innovation programme (Advanced Grant VIRMETAL, grant agreement No. 669141). Support from the Spanish Ministry of Economy and Competitiveness (DPI2015-67667) is also gratefully acknowledged.


\end{document}